\documentstyle[12pt]{article}
\title{{Radiation-Dominated Quantum Friedmann Models}
\thanks{This work is supported in part by funds provided by the U. S.
Department of Energy(D.O.E.) under contract \# DE-FC02-94ER40818 .}}
\author{\sc{Nivaldo A. Lemos}\thanks{On leave of absence from Departamento de
F\'{\i}sica, Universidade Federal Fluminense, Av. Litor\^anea
s/n, Boa Viagem,  24210-340  Niter\'oi, RJ, Brasil.
Supported by Conselho Nacional de Desenvolvimento Cient\'{\i}fico e
Tecnol\'ogico(CNPq), Brasil.}\\
\small{\it Center for Theoretical Physics}\\
\small{\it Laboratory for Nuclear Science
and Department of Physics}\\
\small{\it Massachusetts Institute of Technology}\\
\small{\it Cambridge, MA 02139 - USA}\\
\small{e-mail: lemos@mitlns.mit.edu}
\date{} }

\begin{document}
\pagestyle{myheadings}
\baselineskip 18pt

\maketitle
\begin{abstract}

\vskip .1cm

Radiation-filled Friedmann-Robertson-Walker universes are quantized according
to the Arnowitt-Deser-Misner formalism in the conformal-time gauge.
Unlike previous treatments of this problem, here
both closed and open models are studied, only square-integrable wave
functions are allowed,  and the boundary
conditions to ensure self-adjointness of the Hamiltonian operator are
consistent with the space of admissible wave functions.
It turns out that  the tunneling
boundary condition on the universal wave function
is in conflict with self-adjointness of the Hamiltonian. The
evolution of wave packets obeying different boundary conditions is studied
and  it is generally proven that all models are nonsingular.
Given an  initial condition on the probability density
under which the classical
regime prevails, it is found that
a closed  universe is certain to have an infinite radius,  a density
parameter $\Omega = 1$ becoming a prediction of the theory.
Quantum stationary geometries are shown to exist for the closed universe model,
but oscillating coherent states are forbidden by the
boundary conditions  that
enforce self-adjointness of the Hamiltonian operator.
\\
\\
\vskip .3cm
\noindent PACS numbers: 98.80.Hw , 04.60.Gw
\\
\\

\vfill
\noindent CTP \# 2493 \hfill November 1995
\end{abstract}
\newpage
\noindent{\bf 1. INTRODUCTION}
\vspace{0.75cm}

The lack
of a consistent quantum theory of the full
gravitational field and its sources has stimulated the development of
quantum cosmology,
a less complete but more tractable     method  to investigate the influence
of quantum effects on the evolution of the universe.
The primordial universe, when presumably curvatures and densities approach
the Planck scale, is  believed to be the privileged scenario in which
the quantum aspects of gravity
are expected to become important or even dominant.
In its broadest sense, quantum cosmology consists in ``freezing out''
all but a finite number of degrees
of freedom of the the gravitational field plus its
sources (through imposition of symmetry requirements) and then quantizing
the remaining ones.
This procedure, initiated by DeWitt [1],
is expected to provide some general insights on
what an acceptable quantum theory of gravity should be like, although
it cannot be strictly valid and is open to criticism [2]. Such a
line of attack  has
been extensively explored to quantize  Friedmann-Robertson-Walker(FRW)
universes with
varying matter content such as a scalar field [3-6],
radiation [7-9], a spinor field [10], dust
[9,11-14 ] or a Rarita-Schwinger field [15].

The present paper is dedicated to a further study of
the quantum theory of a radiation-filled FRW
universe. Differently from previous investigations of this system
[7-9], here we discuss
both closed and open models, deal only with
normalizable wave functions, and pay full attention to the domain of
self-adjointness of the Hamiltonian operator. We follow the
Arnowitt-Deser-Misner(ADM) genuine
canonical quantization method [16]. In this approach
one has to  solve the constraint equations at the classical level and
go over to a reduced phase space spanned
by independent canonical variables alone, and this process demands a definite
choice of time. Although often leading to complicated and time-dependent
Hamiltonians, this formalism has the great advantage of reducing the problem
to one of standard quantum mechanics, enabling one to make full use of the
powerful theory of linear operators in Hilbert space. In our treatment
the time variable
is chosen as conformal time, as this enormously simplifies the form of the
Hamiltonian operator, making the quantum dynamics exactly soluble. Further
reasons for choosing conformal time are given in [8].

After the ADM reduction of phase space, only one degree of
freedom remains, which is taken
to be the scale factor $R$.  Since $R$ is restricted to positive values,
it becomes necessary to impose boundary conditions on the wave functions
belonging to the domain of the Hamiltonian operator to ensure its
self-adjointness. For the simplest of such boundary conditions
the time evolution of  wave packets of Gaussian type is worked out, and it is
shown in full generality for the first time
that both the closed and open models are nonsingular. It is remarked
that in the context of the ADM quantization  the so-called tunneling
boundary condition on the universal wave function is in conflict with
self-adjointness of the Hamiltonian operator, or, equivalently, with
unitarity of the quantum evolution. This is not an artifact
of the particular quantization scheme adopted here, since the ADM and
Wheeler-DeWitt descriptions are equivalent for the model at hand [7].
An initial condition
such that the probability density is sharply concentrated at $R=0$ and
under which
the classical regime sets in is considered.
In the  closed case, under such
an extreme initial condition the probability of finding
any finite radius for the universe vanishes. Therefore a density
parameter $\Omega=1$ becomes a prediction of the model, without
neither sacrificing the requirement of square-integrability on the wave
functions nor imposing boundary conditions inconsistent with the space of
admissible state vectors .
A physically questionable aspect of the initial condition adopted
is pointed out, however. Stationary quantum geometries
are shown to exist for the closed model, but the existence of oscillating
coherent
states of the geometry is precluded by the boundary conditions required to
enforce self-adjointness of the Hamiltonian operator, a result at variance
with previous findings [9].

The layout of this paper is  as follows. In Section 2 the classical model is
specified and the ADM reduction of phase space [7] is briefly reviewed.
In Section 3 the problem of the necessary boundary conditions
to ensure self-adjointness of the Hamiltonian operator is considered,
and  the respective propagators
are written down when the two simplest of such boundary conditions
are adopted. In Section 4 the
motion of wave packets obeying different boundary conditions is obtained
in closed form, and it is verified that the singularity is avoided in all
cases. A special initial condition under which
the quantum model is forced into the
classical regime is discussed, and the consequences              for the
case of a closed universe  are considered. Stationary quantum geometries
are taken up  in   Section 5, with particular emphasis on how
the restricted domain of self-adjointness of the Hamiltonian operator
influences such states.
Section 6 is devoted to some final comments.

\vspace {1.25cm}

\noindent{\bf 2. DYNAMICS OF THE CLASSICAL MODEL}

\vspace {.75cm}
The line element for a homogeneous and isotropic universe
can be written in the
FRW form (we take $c=1$)

$$ds^2 = g_{\nu\lambda} dx^{\nu} dx^{\lambda} = - N(t)^2 dt^2
+ R(t)^2 {\sigma}_{ij} dx^i dx^j
\,\,\, , \eqno(2.1)$$
\\
\noindent where ${\sigma}_{ij}$ denotes the   metric for a 3-space
of constant curvature
$k= +1, 0$ or $-1$, corresponding to spherical, flat or hyperbolic spacelike
sections, respectively.

The matter content will be taken to be a perfect fluid, and
Schutz's canonical formulation of the dynamics of a relativistic fluid in
interaction with the gravitational field
will be employed [17]. The degrees of freedom ascribed to
the fluid are five scalar potentials $\varphi , \alpha , \beta , \theta , S$
in terms of which the  four-velocity of the fluid is written as

$$U_{\nu} = \frac{1}{\mu} (\varphi _{,\nu} + \alpha \beta _{,\nu}  +
\theta S_{,\nu}) \,\,\, , \eqno(2.2)$$
\\
\noindent where $\mu$ is the specific enthalpy. By means of the normalization
condition

$$  g_{\nu\lambda} U^{\nu} U^{\lambda} = - 1    \eqno(2.3)$$
\\
one can express $\mu$ in terms of the velocity potentials. The action for the
gravitational field plus perfect fluid is

$$ S= \int_M\,d^4x \sqrt{-g}\,\, ^{(4)}R \, + \,
2\int_{\partial M}\,d^3x\sqrt{h}\, h_{ij} K^{ij} \, + \,
\int_M\,d^4x  \sqrt{-g}\,p \,\,\, \eqno(2.4)$$
\\

\noindent in units such that $c=16\pi G=1$. In Eq.(2.4)
$p$ is the pressure of the fluid, $^{(4)}R$ is the scalar
curvature derived from the spacetime metric $g_{\nu\lambda}$, $h_{ij}$ is the
3-metric on the boundary $\partial M$ of the  4-manifold $M$,
and $K^{ij}$ is the second fundamental
form of the boundary [18]. The surface term is necessary in the
path-integral formulation of quantum gravity in order to rid the
Einstein-Hilbert Lagrangian of second-order derivatives. Variations of the
pressure are computed from the first law of thermodynamics.

Compatibility with
the homogeneous spacetime metric is guaranteed
by taking  all of the velocity potentials of the fluid as functions of $t$
only. We shall take $p = (\gamma - 1) \rho$ as  equation of state
for the fluid, where $\gamma$ is a constant and $\rho$ is the fluid's energy
density (we shall eventually put $\gamma = 4/3$).
In the geometry characterized by (2.1) the appropriate
boundary condition for the
action principle is to fix the initial and final hypersurfaces of constant
time. The second fundamental form of the boundary becomes
$K_{ij} = - {\dot{h}}_{ij}/2N$. As described
in its full details in [7], after inserting the metric (2.1) into the
action (2.4), using the equation of state, computing the canonical momenta
and employing the constraint equations to eliminate the pair $(\theta,
p_{\theta})$, what remains is a reduced action in the Hamiltonian form

$$ S_r = \int dt \Bigl \{ {\dot R}p_R + {\dot \varphi}p_{\varphi}
+{\dot S} p_S
- N {\cal H} \Bigr \} \,\,\,  \eqno(2.5)$$
\\
\noindent where an overall factor of the spatial
integral of $(det \, \sigma)^{1/2}$ has been discarded,
since it has no effect on
the equations of motion. The super-Hamiltonian $\cal H$ is given by

$${\cal H} = - \Biggl ( \frac{p_R^2}{24R} + 6kR \Biggr ) +
p_{\varphi}^{\gamma}\, R^{-3(\gamma - 1)}\, e^S \,\,\, . \eqno(2.6)$$
\\
\noindent  The lapse $N$ plays the role of a Lagrange multiplier,
and upon its variation it is found that the super-Hamiltonian $\cal H$
vanishes.
This is a constraint, revealing that the phase-space
contains redundant canonical
variables.

According to the ADM prescription,
in order to perform a bona fide canonical quantization
one must go over to a reduced phase space spanned by independent canonical
variables alone. This can be achieved         by first making a choice of time
and then solving the super-Hamiltonian
constraint equation  ${\cal H} = 0$ for the canonical variable
conjugate to the time chosen in the first step. This ensures that the final
action preserves its canonical form, and the Hamiltonian
in the reduced phase space is identical to the
canonical variable whose Poisson bracket is unity
with whatever was chosen as time, but now expressed as a
function of the remaining independent canonical variables [16].
In the conformal-time gauge $N=R$ (that is, henceforward $t$
denotes conformal time)
and for $\gamma = 4/3$ (radiation)
this procedure leads to the very simple reduced action [7]

$$ S_r = \int dt \Biggl \{ {\dot R}p_R
- \Biggl ( \frac{p_R^2}{24} + 6kR^2 \Biggr ) \Biggr \} \,\,\, . \eqno(2.7)$$

\noindent Only one degree of freedom is left, namely the scale factor  $R$,
and the Hamiltonian in the reduced phase space is

$$H = \frac{p_R^2}{24} + 6kR^2 \,\,\, . \eqno(2.8)$$
\\
Hamilton's equations of motion lead immediately to

$$ {\ddot R} + kR = 0 \,\,\, . \eqno(2.9)$$
\\
\noindent The solution for $R(t)$ can be written as

$$R(t) = R_0  \left\{ \begin{array}{cl}
              \sin t & \mbox{if $k=+1$} \\
              t & \mbox{if $k=0$}\\
              \sinh t & \mbox{if $k=-1$}
              \end{array} \right.       \eqno(2.10)$$
\\
\noindent with a suitable choice for the origin of  conformal time $t$.
The standard cosmic time $\tau$ is related to conformal time by

$$d\tau = R \, dt \,\,\, , \eqno(2.11)$$
\\
\noindent hence

$$ \tau = R_0 \,  \left\{ \begin{array}{cl}
              1 - \cos t & \mbox{if $k=1$} \\
              t^2/2 & \mbox{if $k=0$}\\
              \cosh t - 1 & \mbox{if $k=-1$}
              \end{array} \right.       \eqno(2.12)$$
\\
\noindent with the convention that $\tau = 0$
when $t = 0$. In the spatially flat
case ($k = 0$), for instance, one recovers the usual behavior $R = C \,
{\tau}^{1/2}$ for the scale factor [19]. It is seen
that Hamilton's principle based on the reduced action (2.7)
gives rise to the same equations of motion as those obtained by first
varying the full action (2.4) and then simplifying them through the use of
the spacetime symmetries of homogeneity and isotropy and of the equation
of state $p = \rho /3$ . Such a consistency check
is indispensable if quantization
in minisuperspace is to have any meaning at all.

\vskip 1.25cm

\noindent {\bf 3. QUANTIZATION, SELF-ADJOINTNESS AND BOUNDARY CONDITIONS}

\vskip .75cm

The remarkably simple form of the Hamiltonian (2.8) makes it possible  to
find exact results for the cosmic evolution at the quantum level. The quantum
dynamics is not so straightforward as one might think at first sight because
the    scale factor $R$ is restricted to the domain $R>0$,
so that the
minisuperspace quantization
in the $R-$representation deals only with wave-functions defined on the
half-line $(0,\infty )$. It is well-known that in such circumstances
one usually has to impose  boundary conditions
on the allowed wave functions otherwise  the relevant
operators will not be self-adjoint, the most important of
all operators being  the  Hamiltonian, that  must be
self-adjoint in order  that the time evolution be unitary.
The need to  impose boundary conditions to ensure self-adjointness
has been  long recognized by practitioners of the
ADM formalism as applied to  quantum cosmology
[3,11]. Very recently it has also been seen to have non-trivial cosmological
implications in the Wheeler-DeWitt approach [20]. What does not appear
to have been duly
emphasized is that self-adjointness conditions depend on the set of allowed
state vectors. It has been argued by adherents of the many-worlds
interpretation [8] and of the pilot-wave formulation [21] of quantum theory
that non-normalizable wave functions are unavoidable in quantum cosmology.
Irrespective  of whether their arguments are physically well founded or not,
what we want to
stress here is that the self-adjointness conditions must be consistent with the
point of view adopted, if the results obtained are to be regarded
as trustworthy.

As follows from the substitution  $p_R \rightarrow
-id/dR$, the Hamiltonian operator associated with the classical Hamiltonian
function (2.8) is (we take $\hbar = 1$)

$${\hat H} = -\, \frac{1}{24} \frac{d^2}{dR^2} + 6k R^2 \,\,\,
\eqno(3.1)$$
\\
\noindent defined on the half-line $(0,\infty )$.  The condition
for $\hat H$ to be symmetric (which, in turn, is a necessary condition for
$\hat H$ to be self-adjoint) is

$$({\psi}_1,{\hat H}\psi _2) =  ({\hat H}\psi _1,\psi _2) \eqno(3.2)$$
\\
\noindent or

$$\int_0^{\infty}\psi _1^*(R)\, \frac{d^2\psi _2}{dR^2}\, dR
=\int_0^{\infty}\,\frac{d^2 \psi _1^*}{dR^2} \, \psi _2 (R)\, dR
\,\,\, , \eqno(3.3)$$
\\
\noindent where the asterisk stands for complex conjugate. Integrating
by parts twice this leads to

$$\Bigl ( \psi _1^*\frac{d\psi _2}{dR} - \frac{d \psi _1^*}{dR} \psi _2
\Bigr ) (\infty ) =  \Bigl ( \psi _1^*\frac{d \psi _2}{dR} -
\frac{d \psi _1^*}{dR}\psi _2
\Bigr ) (0)  \,\,\, . \eqno(3.4)$$
\\

If both $\psi$ and its derivative are square-integrable, the left-hand side of
(3.4) vanishes and
we are left with

$$\Bigl ( \psi _1^*\frac{d \psi _2}{dR} -
\frac{d \psi _1^*}{dR} \psi _2
\Bigr ) (0) = 0 \,\,\, . \eqno(3.5)$$
\\
\noindent Then it can be shown [22] that to ensure the validity of
this condition it is  necessary and sufficient that the domain of $\hat H$
be restricted to those wave functions such that

$$\psi ^{\prime}(0) = \alpha \psi (0) \eqno(3.6)$$
\\
\noindent with $\alpha \in (-\infty ,\infty ]$. This generic boundary
condition was explicitly taken into account in [11] and implicitly used
in the simplest cases $\alpha = 0$ and $\alpha = \infty$ in [7],
in which only square-integrable wave functions were considered acceptable.

If the potential is unbounded from below the Hamiltonian may possess
eigenstates $\psi \in L^2(0,\infty )$
such that ${\psi}^{\prime}$ is not square-integrable. In this case, or if
non-normalizable wave functions are allowed, the correct condition for
symmetry of the Hamiltonian is Eq.(3.4), and as such  it
has been recently  employed in [20]. However, in
[8] non-square-integrable wave functions were argued to be necessary in quantum
cosmology but, inconsistently with this point of view, Eq.(3.6) was imposed on
the
allowed wave functions to allegedly enforce self-adjointness of the
Hamiltonian operator.
As a consequence of demanding that the initial wave function  when $t = 0$
be perfectly
localized at the singularity $R=0$, Tipler [8] was led to a universal
wave function of the form

$$\psi (R,t ) = \Biggl [ \frac{3\, i}{4\, L_P\, \sin t} \Biggr ] ^{1/2}
\exp \, \Bigl [ (3\pi /4i)(\cot t ) (R/L_P)^2 \Bigr ] \equiv
 A(t ) \exp [\, iB(t )R^2\, ] \,\,\, ,\eqno(3.7)$$
\\
\noindent where $L_P$ denotes the Planck length, and
which satisfies the boundary
condition (3.6) for $\alpha = 0$. Since this wave function is not
square-integrable, it should obey Eq.(3.4), that is

$$\lim_{R\rightarrow \infty}\,  [\, 4i A^*  B R \, ] = 0 \,\,\, , \eqno(3.8)$$
\\
\noindent which is not  satisfied because both $A(t )$ and $B(t )$ are
different from zero. Therefore, a wave function of the form (3.7)
cannot represent a possible state of the universe. This means that Tipler's
initial condition on the wave function of the universe together with his
imposition of boundary condition (3.6) with $\alpha =0$ are in conflict with
the
requirement that $\hat H$ be a self-adjoint operator, and this renders
the conclusions
of [8] invalid.

In the present work we shall deal only with square integrable wave functions,
so that  the  set of
admissible  states in the $R$-representation is the Hilbert space
 $L^2(0,\infty )$. Therefore, the domain of self-adjointness of
the Hamiltonian operator is restricted to those wave functions that obey (3.6).
For the sake of simplicity, here we shall
address ourselves only  to
the cases $\alpha=0$ and
$\alpha=\infty$, that is, the boundary conditions we
shall be concerned with are

$$  \psi ^{\prime}(0,t) = 0 \eqno(3.9a)$$

\noindent or

$$\psi (0,t) = 0 \,\,\, . \eqno(3.9b) $$
\\

\noindent Both of these conditions refer to what happens to a wave packet
when it hits the singularity $R=0$. The boundary condition (3.9b) was
advocated    by DeWitt         to keep wave packets away from the
singularity, but in general it is not powerful enough to prevent wave
functions from becoming concentrated in the neighborhood of $R=0$ [23].
As a matter of fact, it has been argued [11] that DeWitt's boundary condition
is just not relevant to the issue of quantum gravitational collapse .

The time development of the models  is fully determined once one is in
possession of the propagator or Green's function. Let $G(x,y,t)$ be the
propagator for the problem in the usual Hilbert space $L^2(-\infty , \infty )$.
Then the propagator for the problem in the restricted Hilbert space
$L^2(0,\infty )$ is

$$G^{(a)}(R,R^{\prime},t) = G(R,R^{\prime},t) + G(R,-R^{\prime},t)
\eqno(3.10)$$
\\

\noindent if the boundary condition is (3.9a), or

$$G^{(b)}(R,R^{\prime},t) = G(R,R^{\prime},t) - G(R,-R^{\prime},t)
\eqno(3.11)$$
\\
\noindent if the boundary condition is (3.9b), as noted by several authors
[24,8,25].

The general Green's function for the Hamiltonian (3.1)
on the usual Hilbert space $L^2(-\infty , \infty)$ is

$$G(x,y,t) = \Biggl [ \frac{6\sqrt{k}}{\pi i \sin (\sqrt{k}\, t)}
\Biggr ] ^{1/2}
\exp\Biggl \{ \frac{6i\sqrt{k}}{\sin (\sqrt{k}\, t)}\Bigl [ (x^2 + y^2)
\cos (\sqrt{k}\, t) - 2xy \Bigr ] \Biggr \} \,\,\, , \eqno(3.12)$$
\\
\noindent as one immediately obtains from the expression of the propagator for
the harmonic oscillator [26] by setting $m=12$ and $\omega = \sqrt{k}$.
In the limiting case $k=0$ the well-known free-particle propagator is
regained, whereas for $k=-1$ all one has to do is make use of the simple
formulae $\cos (it)=\cosh t$ and $\sin (it) = i \sinh t$.
The latter case corresponds
to the quantum mechanics of a particle in an inverted oscillator potential,
studied extensively in [27].

In the quantum cosmology \` a la Hartle-Hawking-Vilenkin-Linde an essential
role is played by initial or boundary conditions. One of these is the
so-called tunneling boundary condition [28,29], according to which
the wave function of the universe must consist only of outgoing modes at
singular boundaries of superspace. In our present context this amounts
mathematically to

$$J = \frac{i}{2} \Biggl ( {\psi}^*\frac{\partial\psi}{\partial R} -
\psi\frac{\partial\psi^*}{\partial R} \Biggr )_{R=0} > 0 \,\,\, ,
\eqno(3.13) $$
\\
\noindent where $J$ is the  probability current density. However, from (3.6)
it follows immediately that $J=0$ because $\alpha$ is a real number. One is
forced to conclude that, at least for this minisuperspace model, the
tunneling boundary condition cannot be implemented  because it is
irreconcilable with self-adjointness of the Hamiltonian operator or,
equivalently,  unitarity of the time evolution. Furthermore, since
the ADM and Wheeler-DeWitt descriptions are equivalent for the present model
[7], this difficulty is not an artifact of the particular
quantization scheme adopted here.

\vskip 1.25cm

\noindent {\bf 4. EVOLUTION OF THE QUANTUM MODELS}

\vskip .75cm

We shall now dedicate some paragraphs to the description of the main features
of the dynamical evolution
of our quantum cosmological models. This will be done by first following the
time development of wave packets and then by studying the effect of imposing
a very special initial condition.

\vskip .6cm
\noindent {\large 4.1 Motion of Wave Packets}
\vskip .2cm

Let us start  by working out the dynamical evolution of  representative
initial wave packets.
The first initial state to be considered is the one described
at $t=0$ by the normalized
wave function

$$ \psi^{(a)}_0 (R) = \Biggl ( \frac{8\sigma}{\pi} \Biggr )
^{1/4} e^{-\beta R^2}
\eqno(4.1)$$
\\
\noindent where $\beta = \sigma + ip$ with $p$ real and $\sigma >0$,
corresponding to the boundary
condition (3.9a).

The initial wave function (4.1) is an even function of $R$, so that

$$\psi^{(a)}(R,t) =\int_0^{\infty} G^{(a)}(R,R^{\prime},t)
\psi^{(a)}_0(R^{\prime})
dR^{\prime} $$

$$= \int_0^{\infty} G(R,R^{\prime},t)\psi^{(a)}_0(R^{\prime})dR^{\prime} +
\int_0^{\infty} G(R,-R^{\prime},t)\psi^{(a)}_0(R^{\prime})dR^{\prime}
= \int_{-\infty}^{\infty} G(R,R^{\prime},t)\psi^{(a)}_0(R^{\prime})dR^{\prime}
\,\,\, . \eqno(4.2)$$
\\
\noindent Inserting the propagator (3.12) and the initial wave function (4.1)
into (4.2) and performing the Gaussian
integration one finds

$$\psi^{(a)}(R,t) =   \Bigl ( \frac{8\sigma}{\pi} \Bigr ) ^{1/4}
\Biggl \{ \frac{6\sqrt{k}}{\cos (\sqrt{k} \, t)[\beta \tan
(\sqrt{k}\, t) -6i\sqrt{k}]} \Biggr \} ^{1/2}$$

$$\times \, \exp \Biggl \{ \frac{6i\sqrt{k}}{\tan (\sqrt{k}\, t)}
\Biggl ( 1+ \frac{6i\sqrt{k}}{\cos ^2(\sqrt{k}\, t)[\beta \tan (\sqrt{k}\, t) -
6i\sqrt{k}]} \Biggr )\, R^2 \Biggr \}\,\,\, . \eqno(4.3) $$
\\

An important quantity is the expectation value of the scale factor

$$\langle {\hat R} \rangle ^{(a)}_t = \int_0^{\infty} R\, \vert \psi^{(a)}
 (R,t) \vert ^2
\, dR \,\,\, , \eqno(4.4)$$
\\
\noindent which can be readily computed from the wave function (4.3). We find


$$ \langle {\hat R} \rangle ^{(a)}_t =  \frac{1}{12}\sqrt{\frac{2}{\pi\sigma}}
   \left\{ \begin{array}{ll}
   \sqrt{\sigma^2 \sin ^2 t + (6-p \tan t)^2 \cos ^2 t} & \mbox{if $k=+1$} \\
\\
     \sqrt{\sigma ^2 t^2 + (6- p t)^2} & \mbox{if $k=0$}\\
\\
     \sqrt{\sigma ^2 \sinh ^2 t + (6-p \tanh t)^2 \cosh ^2 t} & \mbox{if
$k=-1$}
     \end{array} \right.  \,\,\, .     \eqno(4.5)$$

\noindent Notice that, in all cases, $\langle {\hat R} \rangle^{(a)} _t$
never vanishes. For $k=0$ or $k=-1$  and $p>0$ the expectation
value of the scale factor
initially decreases, reaches a minimum value and then grows steadily without
limit, whereas if $p<0$ there is a continuous expansion without bound.
As expected, for large $t$ the highest expansion rate belongs to the
hyperbolic model ($k=-1$). For
$k=1$ the universe oscillates between a minimum and a maximum radius. An
interesting interpretation of this behavior  in terms of
reflection of parts of the wave packet as they hit the origin $R=0$ can be
found in [11].

It should be stressed that the previous results establish that,
at the quantum level, the
singularity is avoided in all cases (that is, $k=0,\pm 1$) according
to the following reasonable  criterion  [11,30]: the
quantum system is singular at a certain instant if $\langle \psi \vert
{\hat f} \vert \psi \rangle = 0$ for any quantum observable $\hat f$ whose
classical counterpart $f$ vanishes at the classical singularity, $\psi$ being
any state of the system at the instant under consideration. For FRW models the
relevant quantum observable is ${\hat f} = \hat R$, since $R=0$
defines the classical singularity. This criterion is in consonance with the
usage in quantum cosmology. Indeed, since $\hat R$ is a positive operator on
$L^2(0,\infty )$, if $\langle {\hat R} \rangle _t = 0$ then $\psi (t)$ is
sharply peaked at $R=0$, and a strong peak in the wave function at a certain
classical configuration is regarded in quantum cosmology as a prediction of
the occurrence of such a configuration [29].

We now turn our attention to the boundary condition (3.9b).
As initial wave function
let us choose

$$\psi _0^{(b)}(R) = \Biggl ( \frac{128\sigma ^3}{\pi} \Biggr ) ^{1/4}
R \, e^{ -\beta R^2} \,\,\, . \eqno(4.13)$$
\\
\noindent Taking advantage of the odd character of this wave function, we can
write

$$\psi^{(b)}(R,t) = \int_0^{\infty} G^{(b)}(R,R^{\prime},t)
\psi_0^{(b)}(R^{\prime})dR^{\prime}
= \int_{-\infty}^{\infty} G(R,R^{\prime},t)\psi_0^{(b)}(R^{\prime})dR^{\prime}
\,\,\, . \eqno(4.14)$$

\noindent Insertion of (4.13) and  (3.12) into  (4.14)
yields

$$ \psi^{(b)}(R,t) =  \Bigl ( \frac{128\sigma ^3}{\pi} \Bigr ) ^{1/4}
\Biggl [ \frac{216 \, i \, k^{3/2}}{\sin ^3({\sqrt k}\, t)} \Biggr ] ^{1/2}
\Biggl [ \beta - \frac{6 i \sqrt{k}}{\tan (\sqrt{k}\, t)} \Biggr ] ^{-3/2}$$

$$\times \, R \, \exp \Biggl \{ \frac{6i\sqrt{k}}{\tan (\sqrt{k}\, t)}
\Biggl ( 1+ \frac{6i\sqrt{k}}{\cos ^2(\sqrt{k}\, t)[\beta \tan (\sqrt{k}\, t) -
6i\sqrt{k}]} \Biggr ) \, R^2 \Biggr \} \,\,\, .\eqno(4.15)$$
\\

\noindent The expectation value
of the scale factor is
found to be


$$\langle {\hat R} \rangle _t^{(b)} = 2\,\langle {\hat R} \rangle _t^{(a)}
\,\,\, , \eqno(4.16)$$
\\
\noindent so that there is no singularity for boundary condition (3.9b) either.

As a matter of fact, we have shown only that the states evolving from (4.1) or
(4.13) are such that $\langle {\hat R} \rangle _t$ never vanishes.
Incomplete analyses like ours of the quantum gravitational collapse problem
have been made before [7,9]. This is insufficient, however, because
in order to
establish that the       quantum cosmological models are nonsingular one has to
prove that $\langle {\hat R} \rangle _t \neq 0$ for {\it any} evolving state
$\psi (t)$ for which  $\langle {\hat R} \rangle _t $ is defined.
A somewhat indirect proof of this will be given below.

Classically the presence of the singularity at $t=0$ makes it physically
mandatory to restrict the conformal time $t$ to positive values. The absence
of singularity at the quantum level makes such a restriction unnecessary, so
that $-\infty < t < \infty$            and the quantum cosmological models
are not naturally endowed with an origin of time. This kind of situation is
also encountered                       in dust-filled FRW models in the
cosmic-time gauge [11].



\vskip .75cm
\noindent {\large 4.2 Special Initial Condition }
\vskip .2cm

Having a quantum dynamical framework to describe the evolution of the universe
is not enough to explain its present state, one has to face the problem of
initial conditions, the gist of modern quantum cosmology.
 In the path-integral       approach to quantum cosmology
both the Hartle-Hawking and the Vilenkin-Linde proposals appear to suffer from
vagueness and  lack of generality, and
can  hardly be said to lead unambiguously to a
unique universal wave function [29]. In the present minisuperspace model the
Vilenkin-Linde tunneling boundary condition is not even implementable, as we
have seen in Section 3.

With this in mind, we proceed tentatively to examine the outcome of imposing
initial
conditions on the probability density associated with the universal wave
function. We content ourselves with discussing  the extreme situation
$\sigma \rightarrow \infty$, in which case
$\vert\psi _0\vert ^2$ becomes sharply concentrated around $R=0$ :

$$ \lim_{\sigma\rightarrow\infty} \vert \psi_0^{(a)}(R) \vert ^2 = \delta (R)
\,\,\, . \eqno(4.17)$$
\\
\noindent Under such circumstances the universe starts with certainty from
 the singularity $R=0$, so that (4.17) may be regarded as the condition for
a quantum explosive birth of the universe, or what might be called
 a quantum big bang.
For $\sigma $ sufficiently large Eq.(4.5) reduces to

$$ \langle {\hat R} \rangle _t \approx  \frac{1}{12}\sqrt{\frac{2\sigma}{\pi}}
              \left\{ \begin{array}{cl}
              \sin t & \mbox{if $k=+1$} \\
              t & \mbox{if $k=0$}\\
              \sinh t & \mbox{if $k=-1$}
              \end{array} \right.       \eqno(4.18)$$
\\
\\
\noindent so that the classical regime sets in $-$ compare  the above
equation with (2.10).

Now for the promised proof that $\langle {\hat R} \rangle _t \neq 0$
for any evolving state $\psi (t)$, implying that our quantum cosmological
models are nonsingular. If $\langle {\hat R} \rangle _{t_1} = 0$
for some $t_1$ then
$\vert \psi (R,t_1) \vert ^2 = \delta (R)$. Suppose   that
$\psi (R,t)$ with $t > t_1$ is an state evolved from
$\psi (R,t_1) $ taken as initial condition. By letting
$\sigma \rightarrow \infty$ it follows from (4.18) that
$\langle {\hat R} \rangle _t = \infty$.
Therefore, no state with finite expectation
value of the scale factor can arise from $\psi (R,t_1) $. Since quantum
mechanics is time reversible, no state $\psi (R,t_0) $ with $t_0 < t_1$
and finite $\langle {\hat R} \rangle _{t_0}$ can evolve
to $\psi (R,t_1) $, which proves that
$\langle {\hat R} \rangle _t \neq 0$   for all evolving
states $\psi (t)$ for which
the expectation value of the scale factor is finite.

Let us focus our attention particularly on the closed
model (k=+1), the only one for which the following considerations
are meaningful.
In our present treatment, that deals only with normalized wave functions,
the probability $P(R<R_1;t)$ that at time $t$ the radius of the universe is
smaller than a given radius $R_1$ is given by

$$P(R<R_1;t)^{(a)} = \int_0^{R_1 } \vert \psi ^{(a)}(R,t) \vert ^2 \, dR =
\Bigl ( \frac{8\sigma}{\pi} \Bigr ) ^{1/2}
\Biggl [ \frac{36}{\cos ^2 t [\sigma ^2 \tan ^2 t +(6 - p\tan t)^2 ]}
\Biggr ] ^{1/2}$$
\\
$$ \times \int_0^{R_1} \exp \Biggl \{ -
\frac{72\sigma R^2} {\cos ^2 t [\sigma ^2 \tan ^2 t +(6 - p\tan t)^2 ]}
\Biggr \}\, dR \,\,\, . \eqno(4.19)$$

\noindent With the change of variable
$$ x = \frac{\sqrt{72\sigma}} {\cos t [\sigma ^2 \tan ^2 t +
(6 - p\tan t)^2 ]^{1/2}} \, R
 \,\,\, , \eqno(4.20)$$
\\
\noindent one gets

$$P(R<R_1;t)^{(a)} = \frac{2}{\sqrt\pi} \int_0^{R_*} e^{-x^2} dx \eqno(4.21)$$
\\
\noindent where

$$R_* = \frac{\sqrt{72\sigma}} {\cos t [\sigma ^2 \tan ^2 t +
(6 - p\tan t)^2 ]^{1/2}} \, R_1   \,\,\, . \eqno(4.22)$$
\\
\noindent Notice that $R_* \rightarrow 0 $ as $\sigma \rightarrow \infty$ ,
hence $P(R<R_1;t) = 0$ for an explosive quantum beginning of the universe.
This leads to the prediction that the density
parameter  $\Omega $ equals unity in the limit of a truly explosive birth
of the universe.
Such a prediction was called ``inflation without inflation'' by Tipler [8],
but here it is derived without having to resort to non-normalizable wave
functions.


If the initial wave function is chosen
as (4.13), corresponding to the boundary condition (3.9b),
Eq.(4.16 ) shows that if $\sigma$ is sufficiently large
the classical regime takes over.
For the closed model one readily finds

$$P(R<R_1;t)^{(b)} = \frac{4}{\sqrt\pi} \int_0^{R_*}x^2\, e^{-x^2}\,
 dx \,\,\, .\eqno(4.23)$$
\\
\noindent Again in the limit $\sigma \rightarrow \infty$ the probability
that at time $t$ the radius of the universe is
smaller than a given radius $R_1$ is zero.
Note that as $\sigma \rightarrow \infty$ the wave packet (4.13) also satisfies
(4.17), thus providing a concrete illustration of the fact that
an initial wave function whose associated probability density is concentrated
entirely at $R=0$ can be harmonized     with the boundary condition (3.9b).
Besides, the prediction $\Omega = 1$ does not appear to be sensitive to
boundary conditions of the type (3.6) on the wave function itself, but
to result exclusively from the  condition (4.17) that its modulus squared
be sharply concentrated at $R=0$ when $t=0$. An initial condition on the wave
function itself that gives rise to $\Omega = 1$ is known [8], but then one
has to give up the square integrability requirement, and the resulting
universal wave function does not belong to the domain of self-adjointness of
the Hamiltonian operator, as remarked in Section 3.

Unfortunately, this state of affairs is still physically dubious.
The inevitable singularity makes the restriction $t\geq 0$ {\it dynamically}
obligatory in classical cosmology. On the other hand, it is only the
{\it imposition} of the initial condition
$\vert \psi _0 (R) \vert ^2 = \delta (R)$ that makes the  instant
$t=0$ so especially distinguished as to induce the restriction $t\geq 0$
upon the quantized model too. This is so because no
unitary evolution from an earlier time        could have led to such a
perfectly localized state at $t=0$. This questionable feature is also present
in Tipler's treatment [8], in which use is made of non-normalizable wave
functions.
Although inconclusive, our tentative considerations were intended to suggest
that it may
be physically reasonable to impose initial conditions on some probability
distribution engendered by the wave function   rather than on the universal
wave function itself.

\vskip 1.25cm

\noindent {\bf 5. QUANTUM STATIONARY GEOMETRIES}

\vskip .75cm

In the case of the closed FRW model the Hamiltonian operator (3.1) possesses
square-integrable eigenfunctions. Consider the normalized wave functions

$$\psi _n(R) = \Biggl ( \frac{\sqrt{48}}{\sqrt{\pi}\, 2^n\, n!} \Biggr ) ^{1/2}
\, H_n\, (\sqrt{12}\, R) \, \exp (-6 R^2) \,\,\,  \eqno(5.1)$$
\\
\noindent where $H_n$ denotes the $n$-th Hermite polynomial.
They satisfy

$${\hat H} \, \psi _n = (n + 1/2)\, \psi _n \,\,\, \eqno(5.2)$$
\\

\noindent where $n$ is a non-negative integer. For even $n$ the wave functions
(5.1) satisfy both Eq.(5.2) and the boundary condition (3.9a), whereas for
odd $n$ they obey both Eq.(5.2) and the boundary condition (3.9b). The effect
of the boundary conditions is to exclude either even or odd eigenfunctions.
This does not agree with [9], where no boundary conditions are imposed on the
wave functions belonging to the domain of $\hat H$, with the result that all
values for $n$ are allowed.
For an universe in any of the stationary states (5.1) nothing changes with
time. This is a purely quantum effect since radiation-filled FRW universes
do not possess classical static solutions.

A more significant effect of imposing self-adjointness boundary conditions
is the preclusion of oscillating coherent states, that is,
nondispersive Gaussian
wave packets whose center oscillates just like a solution to the classical
equations of motion. From the general form of such wave functions [31]
one recognizes at once  that the general boundary
condition  (3.6) cannot be
satisfied even if the real parameter $\alpha$ is allowed to be time dependent.
It is clear, therefore, that a coherent state such that the classical
solution (2.10) emerges as the expectation value of the scale
factor does not exist,
in contradiction with the findings in [9].

\vskip 1.25cm

\noindent {\bf 6. CONCLUDING REMARKS}

\vskip .75cm

In this paper we  dealt with square-integrable wave functions only,
that is, we limited ourselves to the orthodox
framework of quantum mechanics in Hilbert space. Interpretational controversies
apart, it is in this arena that the requirement of self-adjointness
on the quantum observables is most naturally justified and
easily understood. Accordingly, being careful about
domains of operators becomes a necessity in quantum cosmology, and
in the case of  radiation-filled FRW universes the simplest
boundary conditions required to enforce  self-adjointness of $\hat H$ were
taken into account.
Initial or boundary conditions introduced with the purpose of
selecting a unique wave function
are customarily  unrelated to the former, although on occasion
this has been object of
confusion in the literature. Sometimes, however, these two types of boundary
conditions interfere with each other, as we pointed out in Section 3.

An important distinction should be emphasized
between the classical and quantum
cosmologies discussed in this paper.
An origin of time comes into being {\it dynamically} in classical
gravity due to the inevitable singularity, while no origin of time occurs
naturally in quantum gravity, except if  {\it induced} by a choice
of initial conditions. On the other hand, admitting the hypothesis that a
suitable initial condition exists, it is open to doubt whether it should be
imposed on the wave function itself or on some probability distribution
derived thereof.

Our treatment differs from the one based upon consideration
of conformal fluctuations about a given geometry. Apart from lack of
application of boundary conditions to ensure self-adjointness,
there is  a physically more important difference.
In [9]   fluctuations are discussed
about an {\it
specific} FRW classical solution of Einstein's equations, whereas
the ADM  approach  considers fluctuations that encompass all possible universes
of the FRW kind. Although these standpoints may be
classically undistinguishable, they are not necessarily equivalent
in the quantum realm.
This raises the question what the relation between these approaches is
and which, if any,  is appropriate  from the physical
point of view, a study  we reserve for
the future.

We have circumscribed our analysis to radiation as cause of curvature,
that is, to a matter content consisting of
 a perfect fluid with  polytropic index $\gamma = 4/3$. It
can be shown [32] that in the conformal-time gauge
the form of the classical equation of motion of all
Friedmann models with any $\gamma$-fluid as source can be reduced
to  that of a harmonic oscillator after a suitable change of variables. This
suggests the possibility of extending the previous quantum treatment to
cosmological           models whose matter content is a
perfect fluid with an arbitrary polytropic index.
This is presently under investigation.




\newpage

\centerline{\bf REFERENCES}
\begin{description}

\item{[1]} B. S. DeWitt, Phys. Rev. {\bf 160}, 1113 (1967).

\item{[2]} K. V. Kucha\v{r} and M. P. Ryan Jr., Phys. Rev. {\bf D40}, 3982
           (1989).

\item{[3]} W. F. Blyth and C. J. Isham, Phys. Rev. {\bf D11}, 768 (1975).

\item{[4]} N. A. Lemos, Phys. Rev. {\bf D36}, 2364 (1987).

\item{[5]} X. Hu and Y. L. Wu, Phys. Lett. {\bf A125}, 362 (1987).

\item{[6]} N. A. Lemos, {\it Singularities in a Scalar Field Quantum
           Cosmology}, Phys. Rev. {\bf D} (in press).

\item{[7]} V. G. Lapchinskii and V. A. Rubakov, Theo. Math. Phys. {\bf 33},
           1076 (1977).

\item{[8]} F. J. Tipler, Phys. Rep. {\bf 137}, 231 (1986).

\item{[9]} J. V. Narlikar and T. Padmanabhan, Phys. Rep. {\bf 100}, 151 (1983).

\item{[10]} C. J. Isham and J. Nelson, Phys. Rev. {\bf D10}, 3226 (1974).

\item{[11]} M. J. Gotay and J. Demaret, Phys. Rev. {\bf D28}, 2402 (1983).

\item{[12]} P. S. Joshi and S. S. Joshi, Mod. Phys. Lett. {\bf A2},
            913 (1987).

\item{[13]} N. A. Lemos, Phys. Rev. {\bf D41}, 1358 (1990).

\item{[14]} N. A. Lemos, Class. Quantum Grav. {\bf 8}, 1303 (1991).

\item{[15]} T. Christodoulakis and C. G. Papadopoulos, Phys. Rev. {\bf D38},
            1063 (1988).

\item{[16]} R. Arnowitt, S. Deser and C. W. Misner in {\it Gravitation, an
            Introduction to Current Research}, ed. L. Witten (Wiley, New
            York, 1962).

\item{[17]} B. F. Schutz, Phys. Rev. {\bf D2}, 2762 (1970); {\bf D4}, 3559
            (1971).

\item{[18]} See, for example, S. W. Hawking in {\it Quantum Gravity and
            Cosmology}, ed. H. Sato and T. Inami (World Scientific,
            Singapore, 1986).

\item{[19]} See, for example, R. M. Wald, {\it General Relativity}
            (University of Chicago Press, Chicago, 1984).

\item{[20]} J. Feinberg and Y. Peleg, Phys. Rev. {\bf D52}, 1988 (1995).

\item{[21]} Y. Shtanov, {\it Pilot Wave Quantum Gravity}, Preprint ITP-95-8E,
           Bogolyubov Institute for Theoretical Physics (March 1995),
           gr-qc/9503005

\item{[22]} J. von Neumann, {\it Mathematical Foundations of Quantum Mechanics}
           (Princeton University Press, Princeton, 1955).

\item{[23]} N. A. Lemos, Rev. Bras. F\'{\i}s. {\bf 18}, 71 (1988).

\item{[24]} T. E. Clark, R. Menicoff and D. H. Sharp, Phys. Rev. {\bf D22},
            3012 (1980).

\item{[25]} E. Farhi and S. Gutmann, Int. J. Mod. Phys. {\bf A5},
            3029 (1990).
\item{[26]} R. P. Feynman and A. R. Hibbs,
            {\it Quantum Mechanics and Path Integrals}
            (McGraw-Hill, New York, 1968), p. 63.

\item{[27]} G. Barton, Ann. Phys. (N. Y.)  {\bf 166}, 322 (1986).

\item{[28]} A. Vilenkin, Phys. Rev. {\bf D37}, 888 (1988).

\item{[29]} J. A.Halliwell in {\it Quantum Cosmology and Baby Universes}, ed.
S.
            Coleman, J. B. Hartle, T. Piran and S. Weinberg (World Scientific,
            Singapore, 1991). This work contains a guide to the literature on
            quantum cosmology and references to the
            seminal work of Hawking, Hartle, Vilenkin and Linde.

\item{[30]} F. Lund, Phys. Rev. {\bf D8}, 3253 (1973); M. J. Gotay and J. A.
            Isenberg, Phys. Rev. {\bf D22}, 235 (1980); J. A. Isenberg and M.
            J. Gotay, Gen. Rel. Grav. {\bf 13}, 301 (1981).

\item{[31]} L. I. Schiff,
            {\it Quantum Mechanics }
            (McGraw-Hill, New York, 1968), 3rd ed.,  pp. 74-76.

\item{[32]} M. J. D. Assad and J. A. S. Lima, Gen. Rel. Grav.
            {\bf 20}, 527 (1988).

\end{description}
\end{document}